\documentstyle[preprint,pre,aps]{revtex}
\begin{document}
\tightenlines
\draft
\preprint{LF6567E}
\title{Phase Transitions of an Oscillator Neural Network\\ 
with a Standard Hebb Learning Rule}
\author{Toru Aonishi}
\address{
Department of Systems and Human Science, 
Graduate School of Engineering Science,
Osaka University,
1-3, Machikaneyama-cho, Toyonaka, Osaka 560, Japan
}
\date{\today}
\maketitle

\begin{abstract}
Studies have been made on the phase transition phenomena of 
an oscillator network model based 
on a standard Hebb learning rule like the Hopfield model.
The relative phase informations---the in-phase and anti-phase,
can be embedded in the network.
By self-consistent signal-to-noise analysis (SCSNA), 
it was found that the storage capacity is given by $\alpha_c = 0.042$,
which is better than that of Cook's model.
However, the retrieval quality is worse.
In addition, an investigation was made into 
an acceleration effect caused by asymmetry of the phase dynamics.
Finally, it was numerically shown that 
the storage capacity can be improved by modifying
the shape of the coupling function. 
\end{abstract}

\narrowtext

\section{Introduction}

Two hypotheses are available on the carrier of information in the brain.
One, the rate coding hypothesis, states that information is represented
by the density of spikes. 
The other, the temporal coding hypothesis, states that 
information is represented by the timing of neuronal firings,
that is, the synchronization of oscillatory neural activities.

Visual information is divided into a number of features, e.g.,
color, form, motion, and so on, which are processed in parallel 
in different areas of the brain. 
von der Malsburg \cite{malsburg} has emphasized the necessity 
of temporal coding to bind these features, 
and have called into question conventional models based on rate coding. 
It should be noted that some physiological phenomena \cite{singer} 
have supported his discussion.
Therefore, this discussion may also increase interest in temporal coding.

One merit of temporal coding is that 
the information processing (e.g., optimization---minimization 
of an energy function)
achieved by temporal coding neurons (i.e., coincidence detectors) 
is faster than that achieved by rate coding neurons 
(i.e., integrators) \cite{hopfield2}.

Recently, therefore, the oscillator neural network
has been attracting the attention of a growing number of researchers.

In this paper, we discuss a class of oscillator neural
networks that store two-value memory patterns into
synchronously oscillating states. 
This class of oscillator neural networks is difficult to study analytically, 
and therefore was not discussed in previous works \cite{cook,okuda}. 
To overcome the difficulty, we employ a powerful method \cite{fukai3} that 
can be applied to various networks, and derive the memory capacity.
Our model is more biologically relevant than previous models \cite{cook,okuda}, since
it is based on a real number synapse which is specified by
a standard Hebb learning rule.

In previous analyses \cite{cook,okuda,aoyagi} and our analysis,  
a coupling function of a phase equation was approximately assumed as 
the sine function for mathematical tractability. 
In a large population of uniformly coupled oscillators 
with scattered natural frequencies (which corresponds to ferromagnetic models),
the behavior of the system is invariant to the shape of 
the coupling function, that is, the critical variance of the natural frequency
causing a phase transition is only dependent on the lowest
frequency components of the coupling function \cite{daido}.
However, the former analyses do not support our analysis,
since properties of the cross-talk noise in our system are different from
those of scattered natural frequencies which break the mutual entrainment.

In this parer, in order to verify effects of the shape of the coupling function
on the performance of a network, we estimate the memory capacity of more realistic
models. In addition, we numerically show that, in the case of an associative memory 
(on the basis of an oscillator network), gaps in the coupling function 
between in-phase and anti-phase improve the performance.

\section{Analysis of an oscillator network
based on a standard Hebb learning rule}

In general, when the coupling is sufficiently weak,
the high-dimensional dynamics of a coupled oscillator system
can be reduced to the following phase equation \cite{ermentrout},
\begin{eqnarray}
\frac{d \phi_i}{dt} = -\sum_{j=1}^N J_{ij} g(\phi_i -\phi_j); 
\hspace{0.5cm} i=1, \cdots, N, \label{eq2:cupr}
\end{eqnarray}
where $N$ is the total number of oscillators, $\phi_i$ is the 
phase of the $i$th oscillator and $J_{ij}$ denotes a synaptic weight.
$g(\phi)$ is a periodic function. 
Figures \ref{phase}(a)(b) show examples of $g(\phi)$ values (sine-like-curves) 
obtained by the numerical calculation of weakly-coupled BVP 
(Bonhoeffer-Van der Pol) oscillators.

For mathematical tractability,
we approximately assume $g(\phi)=\sin(\phi)$.
Then, Eq. \ref{eq2:cupr} is expressed as,  
\begin{eqnarray}
\frac{d \phi_i}{dt} = -\sum_{j=1}^N J_{ij} \sin(\phi_i -\phi_j). \label{eq2:cupa} 
\end{eqnarray}
In Cook's model (a Q-state clock model with Q$\rightarrow\infty$) \cite{cook}, 
$J_{ij}$ is a complex number synapse specified by  
a generalized Hebb learning rule \cite{aoyagi,okuda}, 
which can not be easily achieved in biological implementation.
Cook's model can be regarded as an extension of Hopfield networks to 
``multi-states''. 
Here, $J_{ij}$ is specified by the following learning rule,
\begin{eqnarray}
J_{ij} = \frac{1}{N} \sum_{\mu=1}^p \xi_i^\mu \xi_j^\mu,
\hspace{0.2cm} \xi_i^{\mu} = \exp( i \theta_i^{\mu}),  \label{eq2:hebb}
\end{eqnarray}
where $\{ \theta_i^\mu \}_{i=1, \cdots, N, \mu=1, \cdots, p}$
are phase patterns to be stored in the network,
and are randomly assigned to $0$ or $\pi$
with a probability of $1/2$. 

We define a parameter $\alpha$ (loading rate) such that $\alpha=p/N$.
This learning rule is equivalent to a standard Hebb learning rule
like in the Hopfield model \cite{hopfield1}, which can be easily achieved 
in biological implementation:
$J_{ij}$ increases between simultaneously firing cells; 
otherwise, $J_{ij}$ decreases.

In area CA1 of the hippocampus, stimulation on the in-phase
of the hippocampal theta rhythm includes a long-term
potentiation that can be depontentiated by stimulation on
the anti-phase of the theta rhythm \cite{holscher}.
Thus, oscillator network models with the standard Hebb learning
are simple and biologically plausible, yet have to be analyzed 
mathematically \cite{okuda,fukai1}.

The system (\ref{eq2:cupa}) has the following potential $V$:
\begin{eqnarray}
V = -\frac{1}{2N} \sum_{ij}^N \sum_{\mu}^p \cos\left((\phi_i-\theta_i^\mu)
-(\phi_j-\theta_j^\mu) \right), \label{eq2:poten}
\end{eqnarray}
This potential (\ref{eq2:poten}) is invariant under the transformation 
$\phi_i \rightarrow \phi_i + \beta$ with any $\beta \in {\bf R}$.
In other words, the equilibrium solution is irresistant
to a uniform shift of $\phi_i$ (the neutral stability). 
Hence, the relative phase informations, i.e., the in-phase and anti-phase,
can be embedded in the network.

Here, we study the equilibrium properties of system (\ref{eq2:cupa})
when $p, N \rightarrow \infty$ with fixed $\alpha=p/N$, 
by using self-consistent signal-to-noise analysis (SCSNA)
\cite{fukai3} which is based on the S/N analysis 
to explain the equilibrium properties.
The results of applying SCSNA to the Hopfield model \cite{fukai3}  
and Cook's model \cite{okuda} 
coincide with those of the replica theory \cite{sompolinsky,cook}.
SCSNA is a powerful method, since we can easily study the 
equilibrium properties of analog-neuron networks without
considering the potential, for example, analogue neural networks 
with asymmetric connections \cite{fukai3}.
 
Defining the order parameters $m_c^\mu$ and $m_s^\mu$ (overlap) 
in a large $N$ limit as,
\begin{eqnarray}
m_c^\mu = \frac{1}{N} \sum_i \xi_i^\mu \cos \phi_i, \hspace{0.5cm}
m_s^\mu = \frac{1}{N} \sum_i \xi_i^\mu \sin \phi_i, \label{eq:over}
\end{eqnarray}
we obtain the following equilibrium condition by setting $d\phi_i/dt=0$ 
in Eq. \ref{eq2:cupa},
\begin{eqnarray}
\cos \phi_i &=& X\left(h^i_1, h^i_2 \right)
= h^i_1 \left/ \sqrt{{h^i_1}^2 + {h^i_2}^2} \right. , \label{eq:equi1}\\
\sin \phi_i &=& Y\left(h^i_1, h^i_2 \right)
= h^i_2 \left/ \sqrt{{h^i_1}^2 + {h^i_2}^2} \right. , \label{eq:equi2}\\
h^i_1 &=& \sum_{\mu}^p \xi_i^\mu m_c^\mu, \hspace{0.5cm} h^i_2 = \sum_{\mu}^p \xi_i^\mu m_s^\mu. \label{eq:locf1}
\end{eqnarray}
Note that we have ignored another solution $\cos\phi_i= -X, \sin\phi_i=-Y$ by applying the so-called Maxwell rule \cite{fukai1} of statistical mechanics to find a relevant solution. Assuming $m_c^1, m_s^1=O(1)$ and $m_c^\mu, m_s^\mu = O(1/\sqrt{N})$ for $\mu>1$, that is, $\xi_i^1$ as a condensed pattern, and expanding 
a polynomial around $m_c^\mu$, $m_s^\mu=0$ $(\mu>1)$
to split the cross-talk noise into an effective self-coupling 
and a Gaussian random variable \cite{fukai3,fukai1},
we obtain the following
nine dimensional equations for the order parameters,
\begin{mathletters}
\label{allequations}
\begin{eqnarray}
& &m_c = \left<\left< X(x_1,x_2) \right>\right>, \label{eq:ope1}\\
& &m_s = \left<\left< Y(x_1,x_2) \right>\right>, \label{eq:ope2}\\
& &q_c = \left<\left< X^2(x_1,x_2) \right>\right>, \label{eq:ope3}\\
& &q_s = \left<\left< Y^2(x_1,x_2) \right>\right>, \label{eq:ope4}\\
& &q_{sc} = \left<\left< X(x_1,x_2) Y(x_1,x_2) \right>\right>, \label{eq:ope5}\\
& &C_1 = \frac{1-\Lambda}{\sqrt{\alpha}}\left<\left< (\overline{Q}_1 x_1 + \overline{Q}_3 x_2) X(x_1,x_2) \right>\right>, \label{eq:ope6}\\ 
& &C_2 = \frac{1-\Lambda}{\sqrt{\alpha}}\left<\left< (\overline{Q}_3 x_1 + \overline{Q}_2 x_2) X(x_1,x_2) \right>\right>, \label{eq:ope7}\\
& &S_1 = \frac{1-\Lambda}{\sqrt{\alpha}}\left<\left< (\overline{Q}_1 x_1 + \overline{Q}_3 x_2) Y(x_1,x_2)\right>\right>, \label{eq:ope8}\\
& &S_2 = \frac{1-\Lambda}{\sqrt{\alpha}}\left<\left< (\overline{Q}_3 x_1 + \overline{Q}_2 x_2) Y(x_1,x_2) \right>\right>, \label{eq:ope9}
\end{eqnarray}
\end{mathletters}
where $\left<\left< \cdots \right>\right>$ is taken 
to mean the Gaussian average over $x_1, x_2$,
that is, $\left<\left< \cdots \right>\right> = \int \int Dx_1
Dx_2\cdots$, and the above equations satisfy $C_2 = S_1$ and $q_c + q_s =1$.
Note that gauge transformations were performed on 
variables of the condensed pattern.
The pattern superscripts $1$ of $m_c, m_s$ are omitted for
brevity.
$\Lambda$ and the Gaussian measure $Dx_1 Dx_2$ 
are expressed as follows,
\begin{eqnarray}
\Lambda &=& C_1 + S_2 + C_2 S_1 - C_1 S_2, \label{eq2:opes1}\\
Dx_1 Dx_2 &=& \frac{dx_1 dx_2}{2 \pi \sqrt{{\rm det} Q}} 
\exp\left(-\frac{1}{2} 
[x_1, x_2] 
Q^{-1}
\left[ \begin{array}{c}
x_1 \\
x_2
\end{array}
\right]
\right), \label{eq2:opes2}\\
Q &=& \left[ \begin{array}{cc}
Q_1 & Q_3 \\
Q_3 & Q_2 
\end{array}
\right], \label{eq2:opes2'}\\
Q^{-1} &=& \frac{1}{{\rm det} Q}\left[ \begin{array}{cc}
Q_2 & -Q_3 \\
-Q_3 & Q_1
\end{array}
\right]
= \left[ \begin{array}{cc}
\overline{Q}_1 & \overline{Q}_3 \\
\overline{Q}_3 & \overline{Q}_2 
\end{array}
\right], \label{eq2:opes3}\\
Q_1 &=& (1-S_2)^2 q_c + 2 C_2(1-S_2)q_{sc} + C_2^2 q_s, \label{eq2:opes4} \\
Q_2 &=& S_1^2 q_c + 2 S_1(1-C_1)q_{sc} + (1-C_1)^2 q_s, \label{eq2:opes5} \\
Q_3 &=& S_1(1-S_2) q_c + (1 - S_2 - C_1 + S_1 C_2 + S_2 C_1) q_{sc} \nonumber \\ 
&+& C_2(1-C_1) q_s.  \label{eq2:opes6}
\end{eqnarray}
$\cos \phi = X(x_1,x_2)$ and $\sin \phi = Y(x_1,x_2)$
at equilibrium satisfy the following condition,
\begin{eqnarray}
& &-\left((1-\Lambda) m_c + \sqrt{\alpha} x_1 \right) \sin \phi +
\left( (1-\Lambda) m_s + \sqrt{\alpha} x_2 \right) \cos \phi
\nonumber \\
& &\hspace{0.5cm} = \alpha  \left[ C_2 \sin^2 \phi - S_1 \cos^2 \phi 
+ (C_1 - S_2) \sin \phi \cos \phi \right]. \label{MX}
\end{eqnarray}
In general, Eq. \ref{MX} admits four solutions 
owing to the effective self-coupling terms
corresponding to the right-hand side of Eq. \ref{MX}.
Figure \ref{maxwell} shows intersection points of the left-hand side 
and the right-hand side of Eq. \ref{MX}, where the filled circles and
open circles correspond to stable solutions and unstable
solutions, respectively. 
Here, we find an available solution 
by applying the Maxwell rule \cite{fukai1} in thermodynamics.
As shown in Fig. \ref{maxwell}, there are two enclosed areas $A_1$ and $A_2$,
provided that the left-hand side of Eq. \ref{MX} is to be the upper
boundary and the right-hand side of Eq. \ref{MX} is to be the lower one.
According to the Maxwell rule \cite{fukai1}, 
we must select a stable solution with the larger enclosed area.

The above order parameters are analogous with those of the Hopfield model 
\cite{sompolinsky}.
$q_c$, $q_s$, and $q_{sc}$ correspond to the so-called Edwards-Anderson order
parameter, which measures the local ordering of XY-spins.
$C_1$, $C_2$, $S_1$, and $S_2$ correspond to the susceptibility,
which measures the sensitivity to external fields. 

The above equations for the order parameters
are more complicated than those of Cook's model
(three-dimensional equations) \cite{cook}.
Because patterns stored in Cook's model are uniformly
distributed in $[0,2 \pi]$,
2-dimensional Gaussian noise--Eq. \ref{eq2:opes2} is isotropic,
which enables us to reduce to three-dimensional equations
in terms of the polar coordinate system.
However, in the case of our model,
2-dimensional Gaussian noise--Eq. \ref{eq2:opes2} is anisotropic,
since phase patterns stored in our model
are only assigned to $0$ or $\pi$.
Thus, we can not reduce to simpler equations any more.

Here, we define a new order parameter $m = \sqrt{m_c^2+m_s^2}$
for convenience.
The solid curve in Fig. \ref{SCSNA}(a) shows 
the values of $m$ for various values of loading rate $\alpha$, which were obtained 
by solving the above equations numerically. 
The plots in Fig. \ref{SCSNA}(a) show results obtained
by numerical simulation with $N=2000$.
Note that the solution of the above equations forms
an isotropic manifold as shown in Fig. \ref{SCSNA}(b),
because system (\ref{eq2:cupa}) is invariant to a uniform shift of $\phi_i$.
Fig. \ref{SCSNA}(a) shows a cross-section of this manifold.

The theory is in good agreement with the results of this simulation.
The storage capacity is given by $\alpha_c = 0.042$ and at this point
the overlaps are $m=0.69$. 
The storage capacity of our model is about $1/3$ as large as 
that of the Hopfield model.
In the case of Cook's model, 
$\alpha_c = 0.038$ and at this point, $m=0.899$.
Thus, the storage capacity of our model 
is better than that of Cook's model,
but the retrieval quality is worse.

The dashed line in Fig. \ref{SCSNA}(a) denotes an unstable solution 
of the equations for the order parameters;
$m_c=1$, $m_s=0$, $q_c=1$, $q_s=0$, 
$q_{sc}=0$, $C_1=0$, $C_2=0$, $S_1=0$, and $S_2=1$. 
The solution corresponds to the equilibrium solution 
$\phi_i = {\theta_i}^1$ (perfect memory states) 
from Eq. \ref{eq2:cupa}.
With a large $N$ limit, the equilibrium solution 
$\phi_i = {\theta_i}^1$ is unstable.

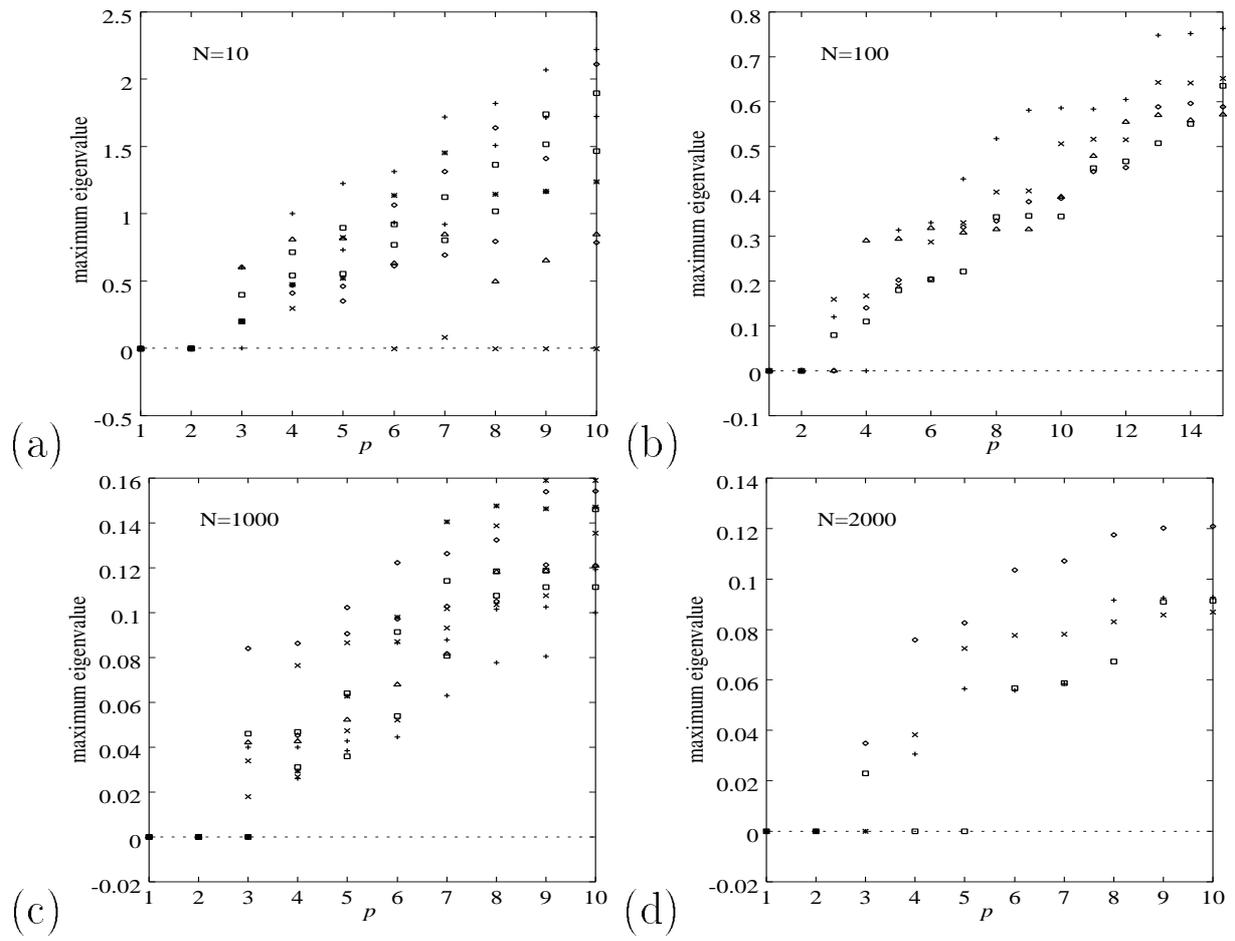
\begin{figure}
\caption{The maximum eigenvalue of matrix $A$ versus the loading rate $\alpha$,
where (a) $N=10$, (b) $N=100$, (c) $N=1000$ and (d) $N=2000$.
At $p\leq2$, the system is neutrally stable around $\phi_i = \theta_i^1$, 
and at $p>2$, the system is unstable, independently of the values of $N$.
}
\label{eigen}
\end{figure}

Let us examine the stability of this solution at a finite $N$.
Linearizing equation (\ref{eq2:cupa}) around $\phi_i = \theta_i^1$, 
we obtain the following matrix $A$:
\begin{eqnarray}
A = \left[ \begin{array}{ccccc}
J_{11}-\sum_{j}^{N} J_{1j} \xi_1^1 \xi_j^1& J_{12}\xi_1^1 \xi_2^1 &
\cdots & J_{1N}\xi_1^1 \xi_N^1\\
J_{21} \xi_2^1 \xi_1^1& J_{22}-\sum_{j}^{N} J_{2j}\xi_2^1 \xi_j^1 & 
\cdots & J_{2N}\xi_2^1 \xi_N^1\\
\vdots & & & \vdots \\
J_{N1}\xi_N^1 \xi_1^1 & J_{N2}\xi_N^1 \xi_2^1 & \cdots &
J_{NN} -\sum_{j}^{N} J_{Nj}\xi_N^1 \xi_j^1
\end{array}
\right],
\end{eqnarray}
where we obtain eigenvalue $0$ with eigenvector 
$\vec{r}=[1, \cdots, 1]^T$ which corresponds 
to a uniform shift of $\phi_i$, 
as previously discussed.
We numerically calculated the maximum eigenvalue of $A$ 
for various values of loading rate $\alpha$.
Figure \ref{eigen} indicates the results, where (a) $N=10$, (b) $N=100$, 
(c) $N=1000$ and (d) $N=2000$.
According to Fig. \ref{eigen}, at $p\leq2$, the system is neutrally stable 
around $\phi_i = \theta_i^1$, and at $p>2$, the system is unstable, 
independently of the values of $N$. 
Thus, we can estimate that the capacity for perfect memory retrieval is given by 
$p_c = 2$.

\section{Acceleration effect}

As shown in Figs. \ref{phase}(a)(c), $g(\phi)$ obtained from real systems 
is asymmetric with respect to the origin, that is, $g(\phi) \neq -g(-\phi)$.
In this case, all phase values $\phi_i$ continue rotating with a uniform frequency,
while keeping the relative states---in-phase and anti-phase.
This phenomenon corresponds to the so-called acceleration effect 
\cite{meunier}.
Let us consider the following perturbed system for Eq. \ref{eq2:cupa},
\begin{eqnarray}
\frac{d \phi_i}{dt} = -\sum_{j=1}^N J_{ij} 
\left(\sin(\phi_i -\phi_j+\sigma)-\sin\sigma\right). \label{eq2:ac1}
\end{eqnarray}
When $\sigma$ is sufficiently small, we obtain the following equation
from Eq. (\ref{eq2:ac1}),
\begin{eqnarray}
\frac{d \phi_i}{dt} = -\sum_{j=1}^N J_{ij} \sin(\phi_i -\phi_j)
- \sigma \sum_{j=1}^N J_{ij} \left(\cos(\phi_i -\phi_j) - 1\right), \label{eq2:ac2}
\end{eqnarray}
where the second term on the right-hand side is a structural
perturbation---asymmetry of $g(\phi)$.
The neutral stability mode (a uniform shift of $\phi_i$) of Eq. \ref{eq2:cupa} 
is broken by this structural perturbation.
Solutions in the neighborhood of an equilibrium solution
for an unperturbed system can be represented as,
\begin{eqnarray}
\phi_i(t)= \psi_i + \omega \tau + \sigma u_i(\tau), \label{eq2:ac3}
\end{eqnarray}
where $\tau=\sigma t$ (slow time variable), and
$\psi_i$ denotes an equilibrium solution for the unperturbed system.
$\omega$ represents a frequency of rotation caused by a structural
perturbation. $\sigma u_i(\tau)$ is a higher-order fluctuation 
caused by the effects
of a perturbation. Substituting Eq. \ref{eq2:ac3} into Eq. \ref{eq2:ac2}, expanding a polynomial around $\sigma=0$ and neglecting higher-order terms, we obtain
\begin{eqnarray}
\omega \left[1, \cdots, 1\right]^T &=& G \left[u_1, \cdots, u_N\right]^T \nonumber \\
&-& \left[\sum_{j=1}^N J_{1j} \left(c_{1j} - 1\right), \cdots, \sum_{j=1}^N J_{Nj} \left(c_{Nj} - 1\right)\right]^T, \label{eq2:ac4}\\
& & \hspace{-3cm} G = \left[ \begin{array}{ccccc}
J_{11} -\sum_{j}^{N} J_{1j}c_{1j} & J_{12}c_{12} & \cdots & J_{1N}c_{1N}\\
J_{21} c_{21} & J_{22} -\sum_{j}^{N} J_{2j}c_{2j} & \cdots & J_{2N}c_{2N}\\
\vdots & & & \vdots \\
J_{N1}c_{N1}& J_{N2}c_{N2} &  \cdots &
J_{NN} -\sum_{j}^{N} J_{Nj}c_{Nj}
\end{array}
\right], \label{eq2:ac5}\\
c_{ij}&=&\cos(\psi_i-\psi_j), \label{eq2:ac6}
\end{eqnarray}
where we obtain $\vec{r}=[1, \cdots, 1]^T \in {\rm ker} G$
which corresponds to a uniform shift of $\phi_i$. 
We can expect ${\rm ker} G = {\rm span} \{\vec{r}\}$,
since, if other modes with eigenvalue $0$ were to exist, 
the relative states would be broken by a perturbation.
We can erase fluctuation term $G \left[u_1, \cdots, u_N\right]^T$ in 
Eq. \ref{eq2:ac4} taking an inner product between $\vec{r}$ and Eq. \ref{eq2:ac4}.
Thus, $\omega$ can be expressed as,
\begin{eqnarray}
\omega = \frac{1}{N}\sum_{ij}^NJ_{ij} 
- \frac{1}{N}\sum_{ij}^N \sum_{\mu}^p \cos\left((\psi_i-\theta_i^\mu)
-(\psi_j-\theta_j^\mu) \right), \label{eq2:ac7}
\end{eqnarray}
where the second term on the right-hand side is the potential $V$ 
of the unperturbed system--Eq. \ref{eq2:cupa}.
Therefore, when $\omega\neq 0$, 
a stationary solution is transformed into a rotating solution 
by a structural perturbation---asymmetry of $g(\phi)$.

\section{Memory capacity of realistic models}

In order to verify effects of the shape of the coupling function
on the memory capacity, we estimated the memory capacity of 
weakly-coupled BVP oscillators with diffusional couplings;
\begin{eqnarray}
& &\left\{ \begin{array}{c}
\tau \dot{x}_i = c \left(x_i - {x_i}^2/3 + y_i \right) + \varepsilon \sum_j J_{ij}
\left(x_j-x_i \right) \\
\tau \dot{y}_i = - \left( x_i + by_i - a \right)/c + \varepsilon \sum_j J_{ij}
\left(y_j-y_i \right)
\end{array} \right. , \label{eq2:bvp}\\ 
& &\hspace{2.5cm} i=1, \cdots, N. \nonumber
\end{eqnarray}
When $\varepsilon$ is sufficiently small,
Eq. \ref{eq2:bvp} can be reduced to the phase-variable description (\ref{eq2:cupr}).
Figures \ref{phase}(a)(b) show $g(\phi)$ values obtained by the 
numerical calculation, where (a) $a=0$, $b=-0.5$, $c=5.0$,  
and (b) $a=0$, $b=0$, $c=5.0$.
We started the numerical calculation using 
phase equation (\ref{eq2:cupr}) with raw $g(\phi)$ values, where N = 1000.

As previously discussed, by the asymmetry of $g(\phi)$,
all phase values $\phi_i$ went on rotating with a uniform frequency,
while keeping the relative states---in-phase and anti-phase.
By using the pattern overlap $m = \sqrt{{m_c}^2 + {m_s}^2}$,
we can  observe the macroscopic relative states, 
since $\left(\sum_i \xi_i \cos(\omega \tau + \phi_i)\right)^2+\left(\sum_i \xi_i \sin(\omega \tau + \phi_i)\right)^2 = \left(\sum_i \xi_i \cos\phi_i \right)^2+\left(\sum_i \xi_i \sin\phi_i\right)^2$. 

Figures \ref{real}(a)(b) show final values of $m$ versus $\alpha$.
The plots were obtained by numerical simulation. 
The solid curves were obtained by SCSNA $( g(\phi)=\sin(\phi) )$.
In Fig. \ref{real}(a), the theory is consistent with the results from the simulation.
In Fig. \ref{real}(b), the memory capacity is larger than the approximated 
system (Eq. \ref{eq2:cupa}). These simulations showed that 
the gaps of $g(\phi)$ in Fig. \ref{phase}(b) improved 
the performance of the network. 

BVP oscillators consist of an activator $x_i$ and an inhibitor $y_i$.
The movement of the activator is rapid like a pulse wave, but
that of the inhibitor is gentle like a sine wave. 
The gaps in $g(\phi)$ in Fig. \ref{phase}(b) were caused by 
a rapid phase-locking of the activator.

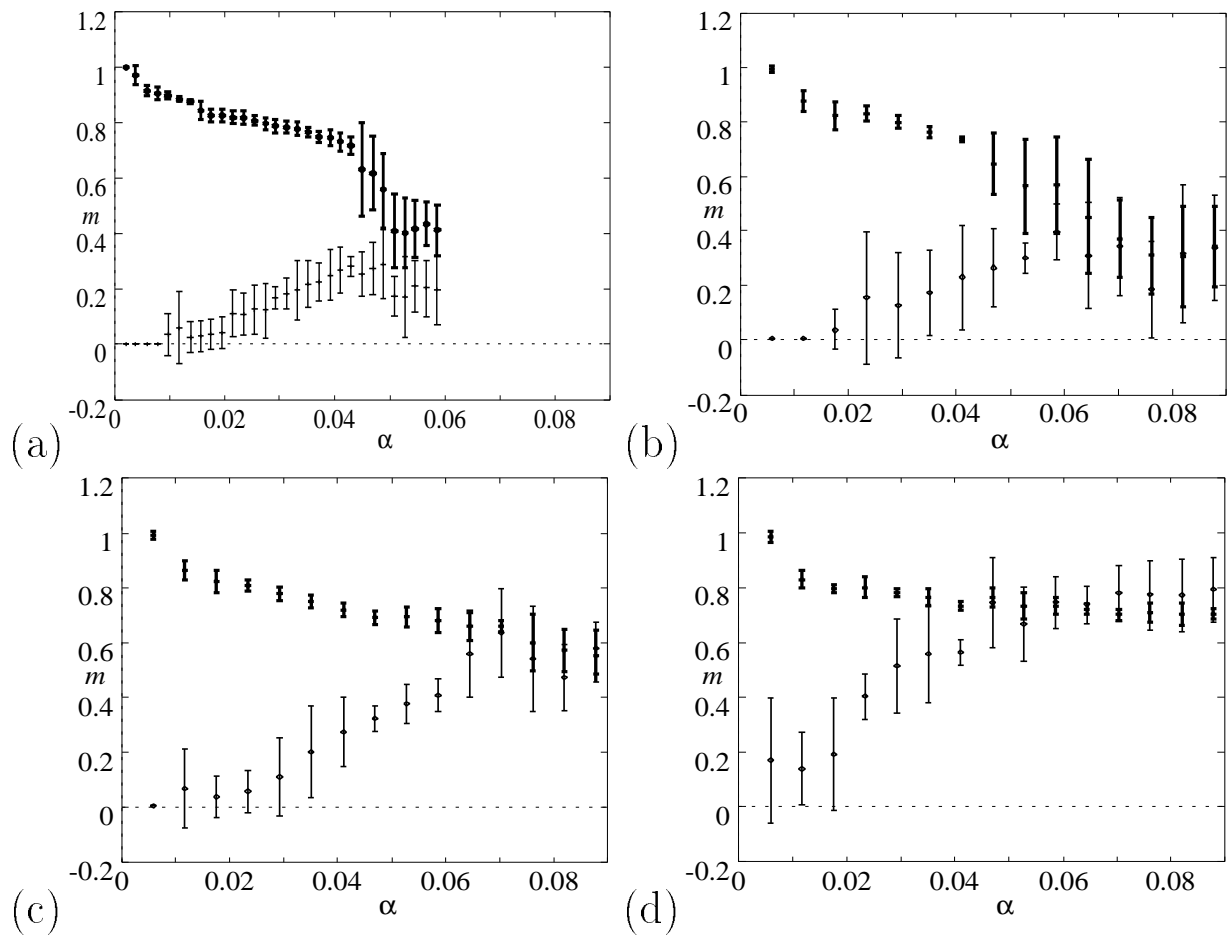
\begin{figure}
\caption{Basin of attraction and memory capacity of an artificial
modal with controllable gaps.
The upper data points represent the equilibrium overlap.
The lower data points denote the boundary of attraction.
All of the data is obtained by numerical simulation 
when $N=512$ for five trials, except Fig. \protect{\ref{ART}(a)} (N=1024). 
The error bars indicate standard deviations. (a) $\gamma=1.0$. (b) $\gamma=0.5$. 
(c)$\gamma=0.25$. (d) $\gamma=0.05$.}
\label{ART}
\end{figure}

We numerically estimated the memory capacity of
the artificial model with the following $g(\phi)$ values,
\begin{eqnarray}
g(\phi) &=& \left\{ \begin{array}{cc}  
\sin(\phi), & -\frac{\pi}{2}+2k\pi <\phi< \frac{\pi}{2}+2k\pi, \\
\gamma \sin(\phi), & {\rm otherwise},
\end{array} \right.\\
& &k = \cdots, -3, -2, -1, 0, 1, 2, 3, \cdots, \nonumber
\end{eqnarray}
where $\gamma$ is the control parameters for the length of the gaps.
Figures \ref{ART}(a)(b)(c)(d) show final values of $m$ and the boundary 
of attraction versus $\alpha$, where (a) $\gamma=1.0$, (b) $\gamma=0.5$, 
$\gamma=0.25$, and (c) $\gamma=0.05$.
In Fig. \ref{ART}(c), the memory capacity is larger than 
that of the original system ($\gamma=1.0$), and the width of the basin 
is almost the same as that in Fig. \ref{ART}(a).
In Fig. \ref{ART}(d), both the memory capacity and the width of the
basin are worse than those in Fig. \ref{ART}(c). 
From these figures, we could guess that the optimal gap is given by
$\gamma \sim 0.25$.

\section{Conclusion}

Studies were made on the phase transitions of 
an oscillator neural network model based 
on a standard Hebb learning rule like in the Hopfield model.
By SCSNA, it was found that the storage capacity is given by $\alpha_c = 0.042$.
In addition, it was numerically shown that 
the gaps in the coupling function between in-phase and anti-phase
improve the performance.
This phenomenon show that a realistic model, i.e., the
pulse-coupled oscillator has a good performance compared to 
a simple ``XY-spin''.

The storage capacity of the Hopfield model can be improved by
replacing the usual monotonic output function 
with a nonmonotonic one \cite{morita}.
The susceptibility becomes negative by the nonmonotonicity
of the output function, so the variance of the cross-talk noise 
decreases. However, under the present conditions, 
we can not determine whether the properties of 
the gaps of the coupling function might be analogous to those of 
the nonmonotonicity in the Hopfield model.
In future work, we need to expand our theory to allow 
it to deal with general $g(\phi)$ values. 

In Cook's model and our model, only replica-symmetric solutions are considered.
SCSNA is based on replica-symmetry, because the Gaussian ansatz for
the cross-talk noise in SCSNA  corresponds to the replica-symmetry ansatz in the replica theory.
It is well-known that replica-symmetry breaking (RSB) 
\cite{almeida} occurs in Cook's model,
since the entropy of this system becomes a negative value \cite{aoyagi3}.
We can also guess that replica-symmetry breaking occurs
in our model. Therefore, we need to derive RSB solutions \cite{parisi}
of the oscillator network model as in the Hopfield model.
However, the derivation of the RSB solutions is more difficult than with 
the Hopfield model.

The author thanks Dr. K. Kurata and Dr. M. Okada
for their valuable discussions.
This work was partially supported by 
Grants-in-Aid for the Encouragement of Young Scientists No. 0782 
and JSPS Research Fellowships for Young Scientists.

\begin{figure}
\caption{(a)(b) Examples of $g(\phi)$ which are obtained by  
the numerical calculation of weakly-coupled BVP oscillators, i.e., Eq. 
\protect{\ref{eq2:bvp}}, where (a) $a=0, b=-0.5, c=5.0$, and (b) $a=0, b=0, c=5.0$. }
\label{phase}
\end{figure}

\begin{figure}
\caption{A schematic diagram of the Maxwell rule to select
relevant solutions to the equilibrium condition.}
\label{maxwell}
\end{figure}

\begin{figure}
\caption{(a) Values of $m$ versus loading rate $\alpha$.  
The solid curve denotes a stable equilibrium
solution, and the dashed line shows an unstable equilibrium solution
which corresponds to $\phi_i = \theta_i^1+\beta$ with any $\beta \in {\bf R}$. 
These lines are obtained by SCSNA.
The plots show results obtained by numerical simulation with $N=2000$.
(b)\ Manifold of the solution. 
Figure \protect{\ref{SCSNA}}(a) shows a cross-section of this
manifold.
}
\label{SCSNA}
\end{figure}

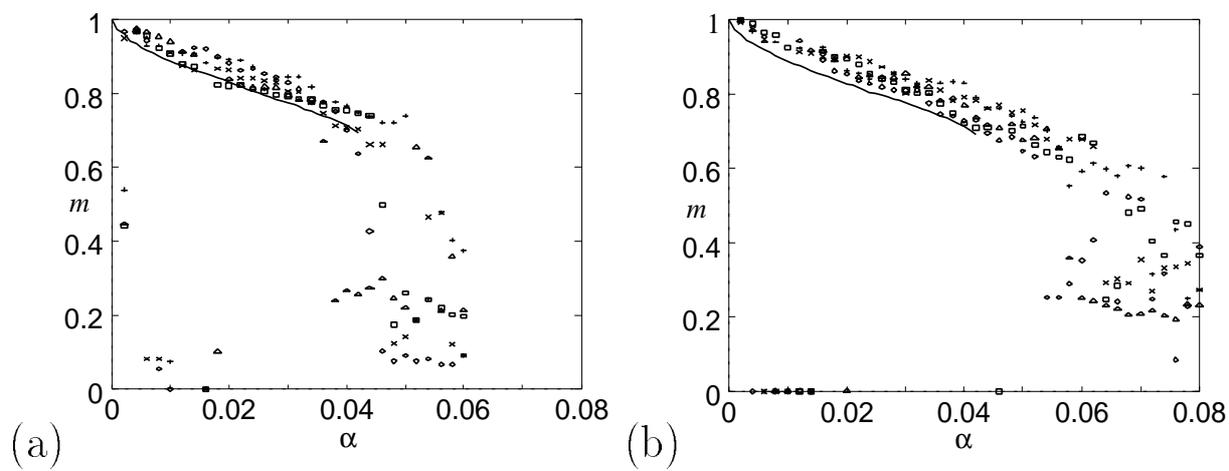
\begin{figure}
\caption{(a)(b) Values of $m$ versus $\alpha$.
The plots are obtained by numerical calculation using 
phase equation (\protect{\ref{eq2:cupr}}) with the $g(\phi)$ values shown in Figs. \protect{\ref{phase}}(a)(b), respectively; N = 1000. 
The solid curves are obtained by SCSNA $( g(\phi)=\sin(\phi) )$.}
\label{real}
\end{figure}

\end{document}